\documentclass[apj]{emulateapj}

\usepackage{times}

\shorttitle{Absolute Abundances in Solar Flares}
\shortauthors{Warren}

\begin{document}

\def\ang{{\small\AA}}
\def\angf{{\scriptsize\AA}}


\title{Measurements of Absolute Abundances in Solar Flares}

\author{Harry P. Warren}

\affiliation{Space Science Division, Naval Research Laboratory, Washington, DC
  20375 USA}


\begin{abstract}
  We present measurements of elemental abundances in solar flares with the EUV Variability
  Experiment (EVE) on the \textit{Solar Dynamics Observatory} (\textit{SDO}). EVE observes both
  high temperature Fe emission lines (\ion{Fe}{15}--\ion{Fe}{24}) and continuum emission from
  thermal bremsstrahlung that is proportional to the abundance of H. By comparing the relative
  intensities of line and continuum emission it is possible to determine the enrichment of the
  flare plasma relative to the composition of the photosphere. This is the first ionization
  potential or FIP bias ($f$). Since thermal bremsstrahlung at EUV wavelengths is relatively
  insensitive to the electron temperature, it is important to account for the distribution of
  electron temperatures in the emitting plasma. We accomplish this by using the observed spectra
  to infer the differential emission measure distribution and FIP bias simultaneously. In each of
  the 21 flares that we analyze we find that the observed composition is close to photospheric.
  The mean FIP bias in our sample is $f=1.27\pm0.23$. This analysis suggests that the bulk of the
  plasma evaporated during a flare comes from deep in the chromosphere, below the region where
  elemental fractionation occurs.
\end{abstract}



\section{introduction}

Solar flares are characterized by the rapid formation of very high temperature and density plasma
in the solar atmosphere. Flares are thought to result from the release of energy from magnetic
reconnection occurring the hot, but relatively tenuous corona \cite[e.g.,][]{priest2002}. This
energy is transported down to the cool, dense chromosphere where it leads to the heating and
evaporation of plasma into the corona \citep[e.g.,][]{fisher1987}.

A similar process has been invoked as a solution to the more general problem of coronal
heating. It has been conjectured that much lower energy magnetic reconnection events (nanoflares,
e.g., \citealt{parker1988}) lead to the formation of the million degree plasma that fills the
upper layers of the solar atmosphere. As in the case of large flares, evaporation plays a central
role in supplying mass to the corona. Intriguingly, the process of bringing mass into the corona
changes its relative composition. The abundance of elements with a low first ionization potential
(FIP $\lesssim 10$\,eV), such as Fe, Si, and Mg, is often enriched in the corona relative to
values measured in the photosphere \citep[e.g.,][]{feldman1992}. The abundance of high FIP
elements, such as O, Ar, and Ne, appears to be unchanged. Thus measurements of plasma composition
hold potential clues as to how mass and energy flow through the solar atmosphere

In this paper we present measurements of elemental abundances observed in solar flares with the
EUV Variability Experiment (EVE, \citealt{woods2012}) on the \textit{Solar Dynamics Observatory}
(\textit{SDO}, \citealt{pesnell2012}). EVE observes a broad range of the solar EUV spectrum
(60--1050\,\ang) at a spectral resolution of about 1\,\ang\ and a cadence of about 10\,s. This
spectral range includes strong emission lines from \ion{Fe}{8} to \ion{Fe}{24} that are formed
over a very wide range of temperatures. This wavelength range also includes continuum emission
from thermal bremsstrahlung \citep{milligan2012} whose intensity is directly related to the
abundance of H. Thus the analysis of EVE spectra can yield measurements of absolute abundance in
flares. To fully account for temperature effects we compute the differential emission measure
distribution using a method described in \citet{warren2013}. Here, however, we consider the line
and continuum contribution to the observed spectra separately and allow for a variable enrichment
relative to the composition of the photosphere. In each of the 21 events that we analyze we find
that the composition is close to photospheric.

\begin{figure*}[t!]
 \centerline{\includegraphics[height=7.0in,angle=90]{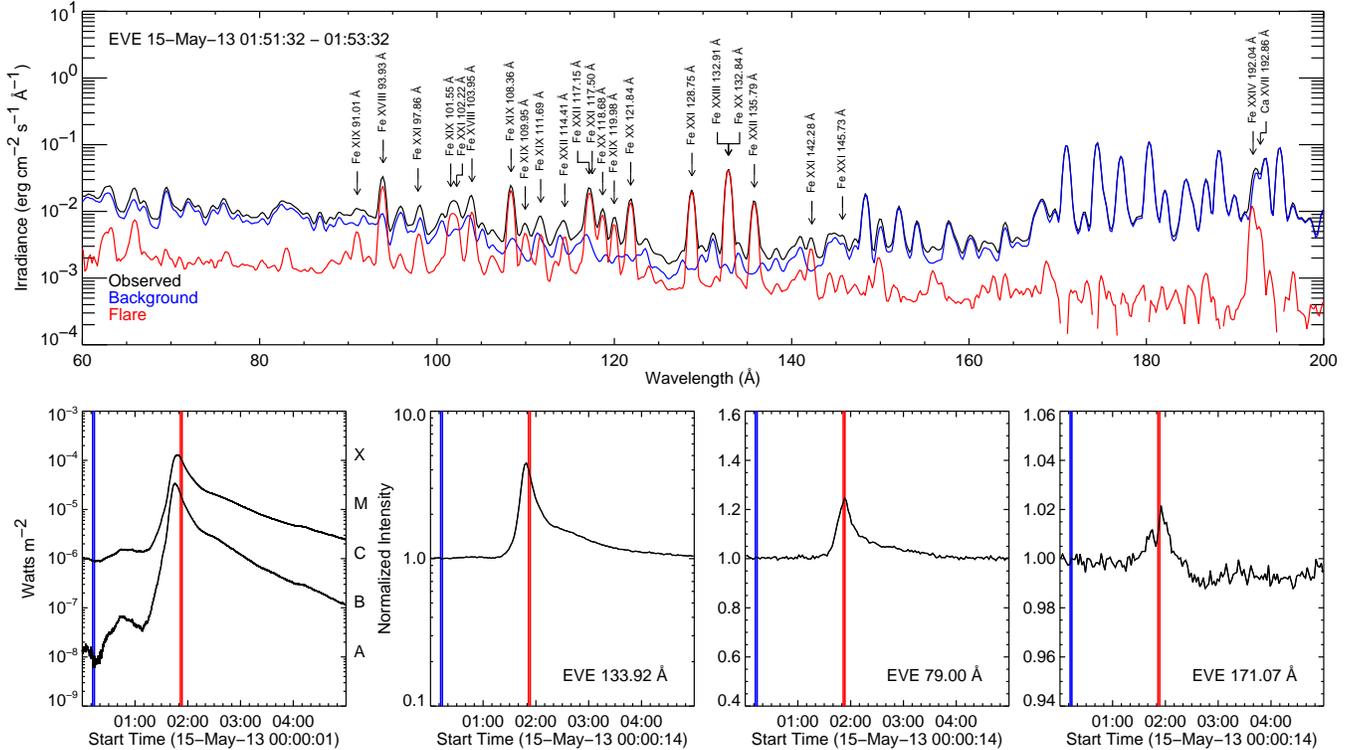}}
 \caption{EVE and GOES observations of the solar irradiance near the peak of an X1.2 flare that
   occurred on 2013 May 15. The top panel shows the EVE spectrum from 60--200\,\angf. The observed
   irradiance, the assumed background irradiance, and the inferred flare irradiance are shown.
   Emission lines from \ion{Fe}{18} to \ion{Fe}{24} and continuum emission are evident. The bottom
   panels show several \textit{GOES} and EVE light curves for this event. The irradiance near
   133\,\angf\ is dominated by \ion{Fe}{23} while the irradiance near 79\,\angf\ is predominately
   continuum. To account for the evolution of the irradiance at lower temperatures and isolate the
   high temperature emission lines during the flare, the background irradiance is assumed to be
   the pre-flare irradiance times the normalized EVE 171\,\angf\ intensity. The red and blue
   vertical lines indicate the time interval used to compute the flare and pre-flare irradiances.}
\label{fig:eve}
\end{figure*}

Past measurements of elemental abundances observed during flares are generally inconsistent with
what we find here. These measurements have often found abundances that are coronal or intermediate
between the coronal and photospheric values \citep{phillips2012,phillips2010,fludra1999,
  fludra1995,schmelz1993a,Schmelz1993b,sterling1993,doschek1985}. EVE observations are unique in
that they cover both a wide range of temperatures and a wide range of wavelengths. As we will
show, our modeled flare spectra largely account for both the wavelength dependence of the
continuum as well as the magnitude of the line emission. We note, however, that some previous work
considered emission lines that are not observed with EVE. \citet{schmelz1993a}, for example,
investigated relative abundances using S and Ne emission lines at soft X-ray wavelengths. Our
results primarily apply to Fe and it may be that the composition during a flare is more
complicated than our analysis suggests or that there is still uncertainty in the photospheric
abundances of the minor ions \citep[e.g.,][]{asplund2009}.

Current models suggest that fractionation in the non-flaring corona occurs at the top of the
chromosphere where the high FIP elements are neutral and low-FIP elements are ionized
\citep[e.g.,][]{laming2004}. The observation of nearly photospheric abundances in solar flares
suggests that the bulk of the plasma evaporated during a flare comes from deep in the
chromosphere. This has implications for how energy is transported from the reconnection region to
the lower layers of the solar atmosphere.

\section{observations}

EVE is actually a collection of instruments designed to measure the solar irradiance at many EUV
wavelengths. In this work we will consider observations from the Multiple EUV Grating Spectrograph
A (MEGS-A), which is a grazing incidence spectrograph that observes in the 50 to 370\,\ang\
wavelength range. MEGS-A has a spectral resolution of approximately 1\,\ang\ and an observing
cadence of 10\,s. For more detail see \cite{woods2012}.

Since EVE observes nearly continuously almost every solar flare is available for analysis. For the
exploratory work considered here we have simply selected the 21 most intense flares for which
there are EVE observations. These events provide ample counts and reduce statistical
uncertainties.

One difficultly with the analysis of EUV spectra at the spectral resolution of EVE is that many of
the emission lines of interest are blended with other emission lines for which there is no
reliable atomic data. This is particularly problematic for the wavelength range between 90 and
150\,\ang\ where there are many unknown emission lines that appear to be formed at temperatures
near 1\,MK \citep[e.g.,][]{testa2012,warren2011}. Given these constraints our strategy is to
remove the lower temperature emission by subtracting a pre-flare observation from the EVE
measurements during the event. The primary risk in this approach is that the lower temperature
emission will also evolve during the flare. For example, in eruptive events dimming is often
observed in emission lines formed around 1\,MK \citep[e.g.,][]{gopalswamy1998}, which leads to a
decrease in the irradiance. Alternatively, bright emission from cooling flare loops will cause the
irradiance from million degree emission lines to increase.

To account for the evolution of the million degree corona during the flare we multiply the
pre-flare spectrum by the normalized irradiance of the \ion{Fe}{9} 171.07\,\ang\ line. An example
EVE irradiance spectrum from the 60--200\,\ang\ wavelength range is shown in
Figure~\ref{fig:eve}. Also shown in Figure~\ref{fig:eve} are EVE light curves for several
wavelength ranges. For this event the 171\,\ang\ light curve indicates a modulation of the million
degree irradiance of approximately $\pm2$\%. For strongest emission lines, such as \ion{Fe}{23}
132.91\,\ang, the contribution of the flare to the irradiance is several times the background
level and such variations in the background are largely irrelevant. For other emission lines the
background makes a more significant contribution to the total emission. The \ion{Fe}{24}
192.04\,\ang\ line is blended with a strong \ion{Fe}{12} line at 192.39\,\ang\ and is particularly
sensitive to background subtraction. 

For this work we have computed time-averaged EVE spectra for each 120\,s interval for which the
\textit{GOES} 1--8\,\ang\ flux is at or above M1 level ($10^{-5}$\,W m$^{-2}$). For the 21 events
under consideration a total of 640 spectra were generated. For each interval we use the observed
standard deviation in the irradiance measurements to estimate the statistical uncertainty in each
spectral bin and propagate the errors to the background subtracted spectra in the usual way.

\begin{figure*}[t!]
 \centerline{\includegraphics[height=6.8in,angle=90]{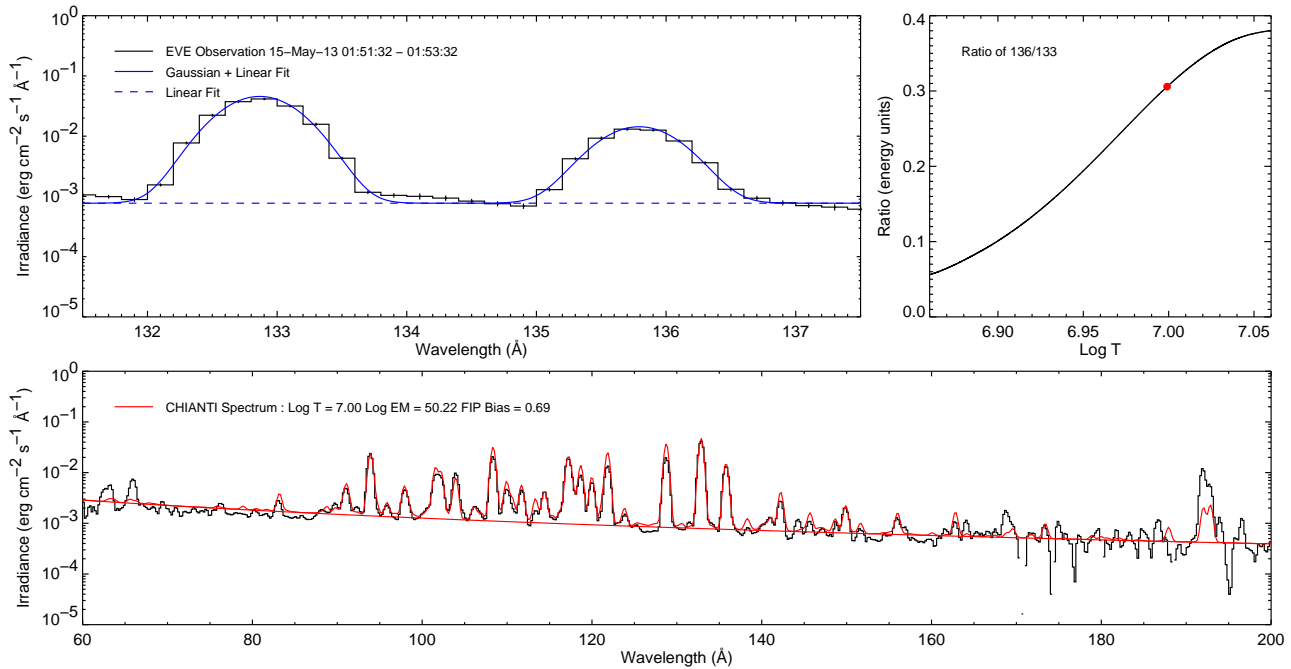}}
 \caption{An isothermal model of the flare emission near 133\,\angf. (\textit{top left panel}) The
   line and continuum intensities are determined by fitting two Gaussians and a constant
   background to the observed emission. The line intensities are used to determine an isothermal
   temperature and emission measure, which are then used to compute the irradiance at these
   wavelengths. The ``FIP bias'' is determined from the ratio of the observed to calculated
   continuum emission near 133\,\angf. For this example the computed FIP bias is 0.69. }
 \label{fig:isothermal}
\end{figure*}

\section{an isothermal model}

In our previous work on the distribution of temperatures in flares observed with EVE we subtracted
the continuum and considered only the contribution of emission lines to the spectrum
\citep{warren2013}. This allowed the shape of the temperature distribution to be determined, but
left some ambiguity as to the magnitude of the distribution. Because of the complexity of emission
measure analysis we first consider a simplified isothermal model that we can use to estimate the
composition of flare plasma. Our previous work has shown that the temperature distribution in a
flare is generally broad and that an isothermal spectrum is often a poor representation of the
observed spectrum. Still, our previous work also suggested that the spectrum between 90 and
150\,\ang\ was often reasonably well approximated by a single temperature model and the simplicity
of this analysis makes it easy to understand. The narrowest temperature distributions are often
observed during the decay of an event and we will focus on this period of the flare. In the next
section we turn to computing the best-fit DEM and the FIP bias simultaneously for both the rise
phase and decay of these flares.

If the solar spectrum in the EUV consisted only of isothermal emission from emission lines and
thermal bremsstrahlung we could model the observations using this simple expression
\begin{equation}
 I(\lambda) = \frac{A}{R^2}\left[f\epsilon_{\rm L}^P(\lambda,T_0){\rm EM_0} + 
      \epsilon_{\rm C}(\lambda,T_0){\rm EM_0}\right],
\label{eq:ints1}
\end{equation}
where $T_0$ and ${\rm EM_0}$ are the isothermal temperature and volume emission measure. The
parameter $A$ is the total area of the flare, $R$ is the Earth-Sun distance, and $A/R^2$ is the
solid angle. The radiated power per unit emission measure (or emissivity) for the emission lines
and continua are $\epsilon_{\rm L}^P$ and $\epsilon_{\rm C}$, respectively. These are calculated
assuming photspheric abundances \citep{grevesse1998}. The factor $f$ is the ``FIP BIAS,'' which
accounts for any deviations from the photospheric composition assumed in the emissivity
calculations. We assume that all of the emission lines of interest are low FIP and that they all
have the same enrichment.

One might imagine that the ideal strategy is to isolate the continuum emission and determine the
magnitude of the emission measure from it. The continuum emission, however, is only weakly
dependant on temperature and there is no unique solution. Instead we use emission line ratios to
infer the isothermal temperature and the product of the FIP bias and the emission measure (${\rm
  EM_0^\prime}=f\cdot{\rm EM_0}$). Then by comparing the observed and modeled continuum emission
we can infer the magnitude of the FIP bias. That is, we rewrite Equation~\ref{eq:ints1} as
\begin{equation}
 I(\lambda) = \frac{A}{R^2}\left[\epsilon_{\rm L}^P(\lambda,T_0){\rm EM_0^\prime} + 
      \epsilon_{\rm C}(\lambda,T_0)\frac{\rm EM_0^\prime}{f}\right].
\label{eq:ints2}
\end{equation}
To further simplify matters we derive the temperature and emission measure from the ratio of the
\ion{Fe}{23} and \ion{Fe}{22} features near 133\,\ang. 

In Figure~\ref{fig:isothermal} we show the application of this analysis to a spectrum from the
2013 May 15 X1.2 flare. The comparison of the observed intensity ratio to the theoretical ratio,
which is computed from the CHIANTI atomic physics database
\citep[e.g.,][]{dere1997,dere2009,landi2012}, yields the temperature and emission measure. Using
these parameters we synthesize the expected continuum emission. The FIP bias parameter is then
determined from the ratio of the computed to observed continuum. Finally, we add the line and
continuum emission together to calculate the expected EVE spectrum. As is shown in
Figure~\ref{fig:isothermal} this procedure yields a reasonable fit to the line and continuum
emission in the 60 to 150\,\ang\ wavelength range for a FIP bias close to 1.

The spectrum shown in Figure~\ref{fig:isothermal} was taken after the peak in the \textit{GOES}
flux when the temperatures are down somewhat from the peak \citep[e.g.,][]{sterling1997} and we
might expect the isothermal approximation to be somewhat more useful.  We have repeated the
isothermal analysis on each of the 21 largest flares observed by EVE. For each event we pick the
first spectrum taken after the temperature derived from the ratio of the \textit{GOES} channels
falls below 15\,MK. Note that the \textit{GOES} temperature plays no role in the analysis of the
EVE observations other than to select the spectrum for analysis.

The result of this calculation for each event is given in Table~\ref{table:fip}. The mean of the
measured FIP bias factors is $f=0.85\pm0.16$, which is close to a photospheric composition. This
approach provides an estimate of the composition using a very simple model.

\begin{deluxetable}{lrrrrr@{$\pm$}r}
\tablewidth{3.1in}
\tablecaption{EVE Abundance Measurements\tablenotemark{a}}
\tablehead{
\multicolumn{2}{c}{} &
\multicolumn{2}{c}{Isothermal} &
\multicolumn{1}{c}{} &
\multicolumn{2}{c}{DEM} \\ 
[.3ex]\cline{3-4}\cline{6-7}\\[-1.6ex]
\multicolumn{1}{c}{Date} &
\multicolumn{1}{c}{Class} &
\multicolumn{1}{c}{$T_{start}$} &
\multicolumn{1}{c}{$f$} &
\multicolumn{1}{c}{} &
\multicolumn{2}{c}{$f\pm\sigma_f$}
}
\startdata
09-Aug-11 & X6.9 & 08:21:10 &     0.97 &  &     1.25 &     0.14 \\
07-Mar-12 & X5.4 & 00:52:14 &     0.73 &  &     1.25 &     0.14 \\
14-May-13 & X3.2 & 01:31:07 &     0.75 &  &     1.25 &     0.14 \\
13-May-13 & X2.8 & 16:27:04 &     0.85 &  &     1.11 &     0.33 \\
15-Feb-11 & X2.2 & 02:13:21 &     0.54 &  &     1.06 &     0.19 \\
06-Sep-11 & X2.1 & 22:35:42 &     0.75 &  &     1.23 &     0.22 \\
03-Nov-11 & X1.9 & 20:34:20 &     1.01 &  &     1.38 &     0.16 \\
24-Sep-11 & X1.9 & 09:55:24 &     0.73 &  &     1.11 &     0.50 \\
23-Oct-12 & X1.8 & 03:27:11 &     0.94 &  &     1.27 &     0.14 \\
07-Sep-11 & X1.8 & 22:47:52 &     0.95 &  &     1.24 &     0.27 \\
13-May-13 & X1.7 & 02:39:30 &     0.89 &  &     1.26 &     0.24 \\
27-Jan-12 & X1.7 & 18:37:41 &     1.33 &  &     1.28 &     0.13 \\
09-Mar-11 & X1.5 & 23:27:19 &     0.98 &  &     1.28 &     0.14 \\
12-Jul-12 & X1.4 & 17:21:12 &     0.72 &  &     1.36 &     0.21 \\
22-Sep-11 & X1.4 & 11:24:33 &     0.85 &  &     1.44 &     0.28 \\
07-Mar-12 & X1.3 & 00:52:14 &     0.73 &  &     1.25 &     0.14 \\
15-May-13 & X1.2 & 01:51:32 &     0.69 &  &     1.26 &     0.12 \\
06-Jul-12 & X1.1 & 23:13:14 &     0.73 &  &     1.19 &     0.17 \\
05-Mar-12 & X1.1 & 04:23:29 &     0.89 &  &     1.28 &     0.11 \\
04-Aug-11 & M9.3 & 03:59:07 &     0.83 &  &     1.32 &     0.15 \\
30-Jul-11 & M9.3 & 02:11:49 &     0.95 &  &     1.37 &     0.12
\enddata
\tablenotetext{a}{FIP bias calculations for the 21 largest flares observed with EVE. For the
  isothermal model only a single spectrum selected during the decay of the event is analyzed. The
  mean FIP bias is $f=0.84\pm0.16$. For the DEM model all of the spectra observed when the
  \textit{GOES} flux is above M1 are considered. A total of 640 spectra have been fit for the DEM
  model and the mean FIP bias for all of the measurements is $f=1.27\pm0.23$.}
\label{table:fip}
\end{deluxetable}

\begin{figure*}[t!]
 \centerline{\includegraphics[height=6.8in,angle=90]{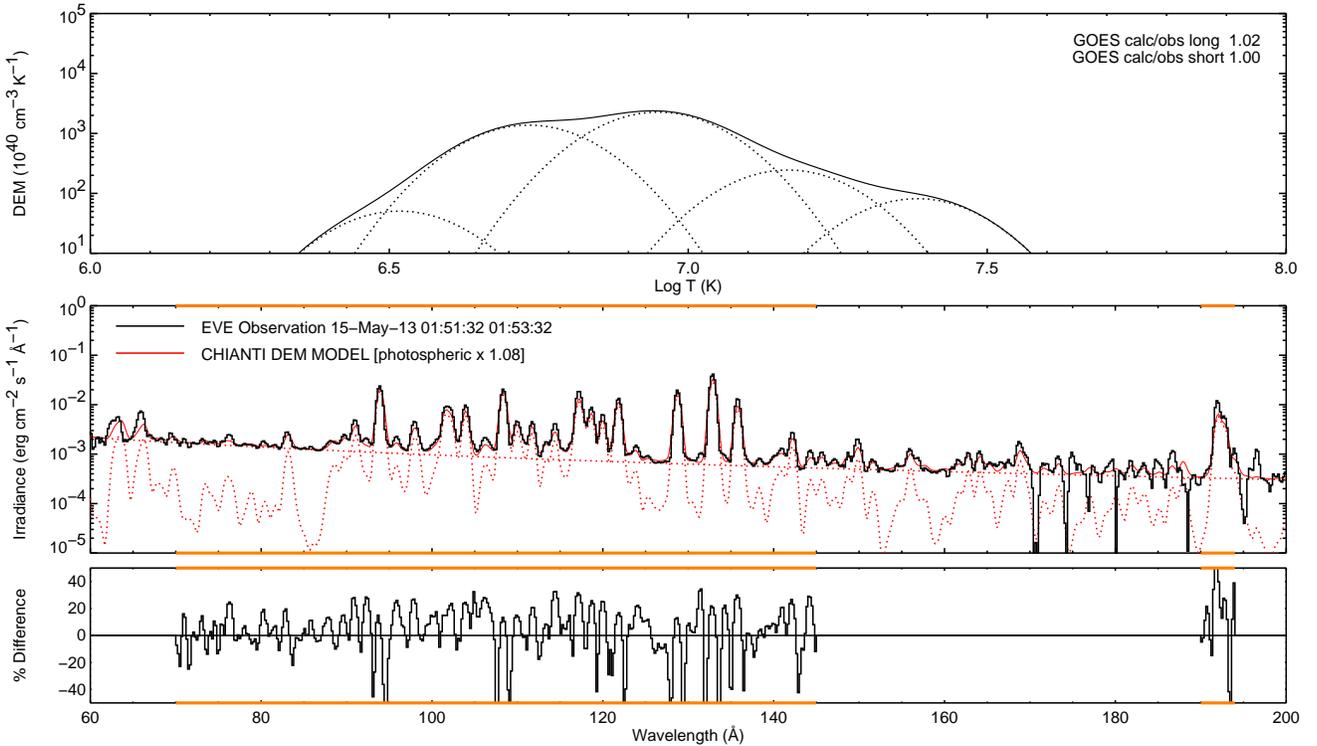}}
 \caption{A differential emission measure analysis of EVE and \textit{GOES} flare
   observations. The top panel shows the best-fit DEM. Individual components of the DEM are
   indicated by the dotted lines. The middle panel shows the observed and modeled spectra. The
   dotted lines show the contribution of the lines and continua to the CHIANTI spectrum. The
   bottom panel shows the difference between the model and the observation. The best-fit FIP bias
   parameter for this observation is $f=1.08$.}
\label{fig:dem}
\end{figure*}

\begin{figure*}[t!]
 \centerline{%
   \includegraphics[width=3.4in]{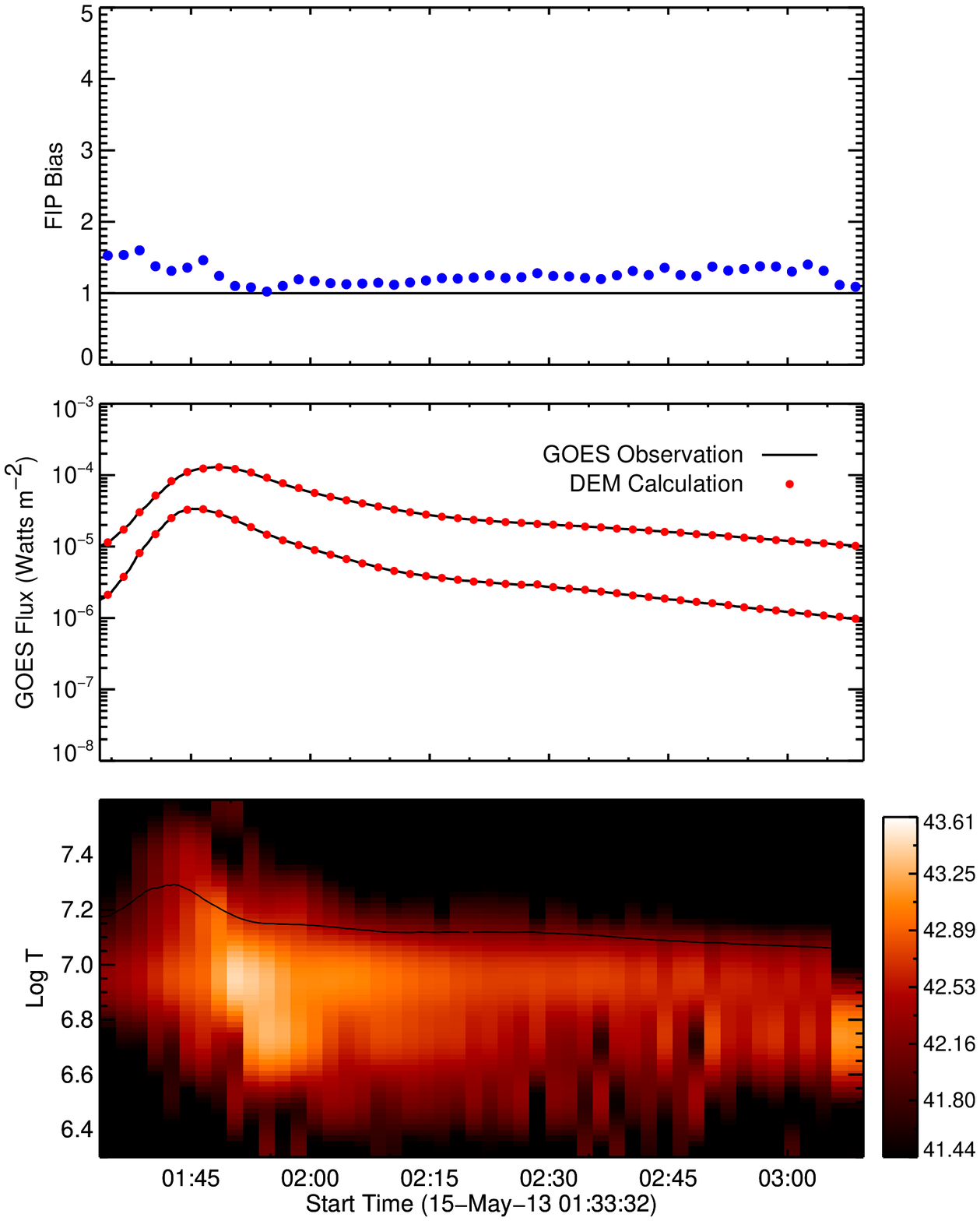}
   \includegraphics[width=3.4in]{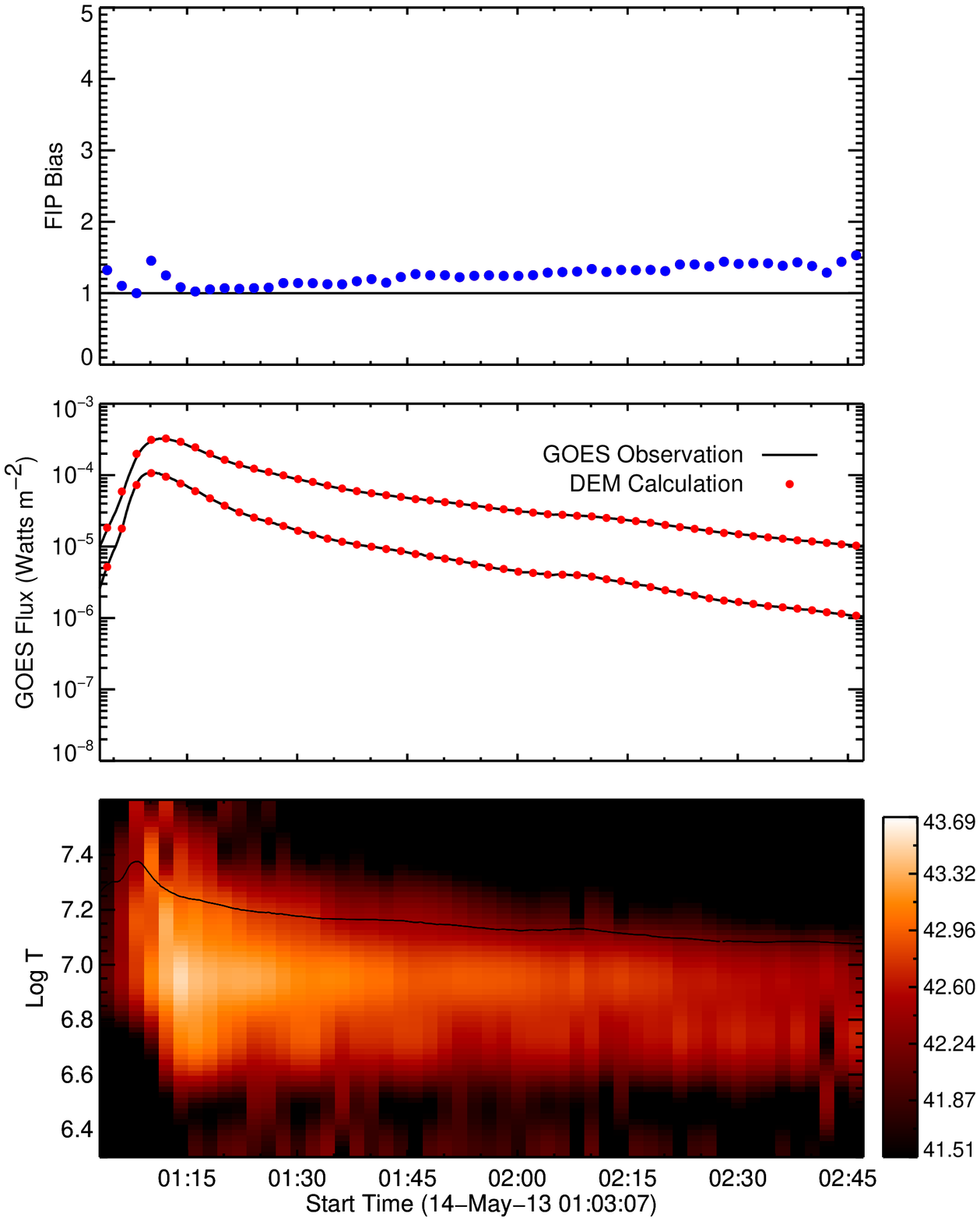}
 }
 \caption{Temporally resolved measurements of elemental abundances for two events observed with
   EVE and \textit{GOES}. The bottom panels show the DEM computed for each event as a function
   of time. The solid black line is the temperature inferred from the ratio of the \textit{GOES}
   soft X-ray channels. The middle panels show the temporal evolution of the observed and computed
   \textit{GOES} soft X-ray irradiances. The top panels show the best-fit FIP bias parameters
   ($f$) as a function of time. }
 \label{fig:fip}
\end{figure*}

\section{a dem model}

To fully account for the distribution of temperatures in the flare we must compute the
differential emission measure or DEM. The DEM represents an empirical description of the solar
atmosphere and is determined by inverting the ill-posed integral equation
\begin{equation}
   I(\lambda) = \frac{A}{R^2}\left[
     \int\left(f\epsilon_L(\lambda,T_e)+\epsilon_C(\lambda,T_e)\right)
             \xi(T_e)\,dT_e\right],
\label{eq:flux}           
\end{equation}
where, as before, $\epsilon_L(\lambda,T_e)$ and $\epsilon_C(\lambda,T_e)$ are the emissivities of
the emission lines and continuua computed with CHIANTI. The function $\xi(T_e)=n_e^2ds/dT$ is the
line of sight DEM. Note that the spatially unresolved EVE observations yield a volume emission
measure ($\xi_V=A\xi(T_e)$) which incorporates the area factor into the line-of-sight emission
measure.

As in our previous work we represent the DEM as a sum of Gaussians in log space
\begin{equation}
   \xi_V(T_e) = \sum_{k=1}^{N_g} \textrm{EM}_k \exp\left[
      -\frac{(\log T_e-\log T_k)^2}{2\sigma_k^2}\right],
\end{equation}
where the number ($N_g$), position ($\log T_k$), and width ($\sigma_k$) of the Gaussians is fixed
for a given calculation and only the magnitude of each component is varied. We select random
initial values for $\textrm{EM}_k$, initialize $f=1$, and use the Levenberg-Marquardt
least-squares minimization routine {\tt MPFIT} \citep{markwardt2009} to determine the values for
the emission measure components and the FIP bias that produce the lowest value of $\chi^2$. The
entire spectrum is not used to compute the deviates, but only spectral regions that contain
strong, optically thin flare emission lines.  The \textit{GOES} fluxes are used as additional
constraints in computing the DEM. They influence the DEM the highest temperatures, but don't play
a major role in determining the shape of the distribution. See \citealt{warren2013} for additional
details on all aspects of the calculation.

An example of this calculation is shown in Figure~\ref{fig:dem}, where the same spectrum
considered for the isothermal fit (Figure~\ref{fig:isothermal}) has been analysed. The peak of the
DEM is near 10\,MK, as one would anticipate from the single temperature model, but the improved
fit to the observed spectra clearly requires significant emission over a broad range of
temperatures. The best-fit value for the FIP bias for this spectrum is $f=1.08$, very close to
photospheric.

As in our previous work we can calculate the best-fit DEM and FIP bias parameters for each
spectrum in each of the 21 flares in our sample. As we stated earlier we have considered
observations for times when the \textit{GOES} long wavelength flux is above M1 and averaged each
spectra over 120\,s intervals. This yields a total of 640 spectra for which we have performed
calculations. Calculations for two representative events are illustrated in Figure~\ref{fig:fip},
which show the DEM, \textit{GOES} flux, and FIP bias as a function of time. Almost all of the
computed FIP bias parameters are close to 1. Only 69 spectra, or about 11\%, indicate a FIP bias
above 1.5. Considering all of the measurements the mean FIP bias is $f=1.27\pm0.23$.

We note two features of the time-dependant FIP bias calculations. There tends to be more scatter
in the measurements during the impulsive phase of the event, perhaps because the temperature
distributions are generally broader and the FIP bias is less well constrained during these
times. We also notice secular trends in the FIP bias, such as is seen in the top panels of
Figure~\ref{fig:fip}, where the parameter increases over time. The leads to an enhancement of the
variance in the measurements. 

\section{summary and discussion}

We have presented measurements of absolute abundances during solar flares observed with the EVE
irradiance instrument on \textit{SDO}. These measurements provide compelling evidence that the
composition is close to photospheric at all times during a flare. Coronal plasma often shows an
enrichment of low FIP elements of about 4 \citep{feldman1992}. The mean FIP bias for the 21 large
flares considered here is $f=1.27\pm0.23$, suggesting only a slight enhancement.

These results do not agree with many previous measurements which have generally indicated an
enrichment of a factor of 2 or more during flares \citep{phillips2012,phillips2010,fludra1999,
  fludra1995,schmelz1993a,Schmelz1993b,sterling1993,doschek1985}. We consider our measurements to
be the most comprehensive ever undertaken. EVE has a broad temperature coverage, which includes
emission lines from \ion{Fe}{15} to \ion{Fe}{24}, continuous observing, and the sensitivity to
observe continuum emission over a very wide wavelength range. The ability of the DEM model to
reproduce the wavelength dependence of the continuum emission from 60 to 200\,\ang\ is the most
compelling aspect of this analysis. We again note, however, that some previous work considered
emission lines that are not observed with EVE. \citet{schmelz1993a}, for example, investigated
relative abundances using S and Ne emission lines at soft X-ray wavelengths. Our results primarily
apply to Fe and it may be that the composition during a flare is more complicated than our
analysis suggests or the measured photospheric composition for these elements is more uncertain
\citep[e.g.,][]{asplund2009}.

Our results suggest that the bulk of the flare is plasma evaporated from deep in the chromosphere,
below the layer at which fractionation occurs, and the \textit{in situ} heating of coronal plasma
does not make a significant contribution to the observed emission. This result needs to be
reconciled with simulations of both the FIP effect \citep[e.g.,][]{laming2004} and chromospheric
evaporation. These results also have implications for simulations of the heating and cooling of
flare plasma \cite[e.g.,][]{warren2005}.


\acknowledgments The \textit{SDO} mission and this research was supported by NASA. CHIANTI is a
collaborative project involving Naval Research Laboratory (USA), the Universities of Florence
(Italy) and Cambridge (UK), and George Mason University (USA). The author benefited greatly from
discussions of coronal dimming with members of the Coronal Dimming Working Group at the 2013 EVE
Science Team Meeting held in Boulder, Colorado.


\bibliography{apj}

\end{document}